\documentclass[journal,12pt,onecolumn,draftclsnofoot,]{IEEEtran}
\IEEEoverridecommandlockouts
\usepackage{float}
\usepackage[table]{xcolor}
\usepackage{cite}
\usepackage{subcaption}
\usepackage{multirow}
\usepackage{graphicx}
\usepackage{amsmath,amssymb,amsfonts}
\usepackage{algorithmic}
\usepackage{graphicx}
\usepackage{textcomp}
\usepackage{xcolor}
\usepackage{multirow}
\usepackage{tabularx}
\usepackage{url}
\usepackage[super]{nth}

\def\BibTeX{{\rm B\kern-.05em{\sc i\kern-.025em b}\kern-.08em
    T\kern-.1667em\lower.7ex\hbox{E}\kern-.125emX}}

\begin{document}

\title{Toward Explainable Users: Using NLP to Enable AI to Understand Users’ Perceptions of Cyber Attacks}

\author{\IEEEauthorblockN{Faranak Abri, Luis Felipe Gutiérrez, Chaitra T. Kulkarni, Akbar Siami Namin, and Keith S. Jones}
\IEEEauthorblockA{ \textit{Texas Tech University}\\
Lubbock, Texas, USA \\
\{faranak.abri, luis.gutierrez-Espinoza, chaitra-tirumalesh.kulkarni, akbar.namin, keith.s.jones\}@ttu.edu}
}

\maketitle

\begin{abstract}
To understand how end-users conceptualize consequences of cyber security attacks, we performed a card sorting study, a well-known technique in Cognitive Sciences, where participants were free to group the given consequences of chosen cyber attacks into as many categories as they wished using rationales they see fit. The results of the open card sorting study showed a large amount of inter-participant variation making the research team wonder how the consequences of security attacks were comprehended by the participants. As an exploration of whether it is possible to explain user's mental model and behavior through Artificial Intelligence (AI) techniques, the research team compared the card sorting data with the outputs of a number of Natural Language Processing (NLP) techniques with the goal of understanding how participants perceived and interpreted the consequences of cyber attacks written in natural languages. The results of the NLP-based exploration methods revealed an interesting observation implying that participants had mostly employed checking individual keywords in each sentence to group cyber attack consequences together and less considered the semantics behind the description of consequences of cyber attacks. The results reported in this paper are seemingly useful and important for cyber attacks comprehension from user's perspectives. To the best of our knowledge, this paper is the first introducing the use of AI techniques in explaining and modeling users' behavior and their perceptions about a context. The novel idea introduced here is about explaining users using AI.  
\end{abstract}


\section{Introduction}

One of the objectives of an ongoing project 
is to design a user-centric threat model by which a cyber attack is described in terms of its immediate non-technical consequences to end-users (i.e., an end-user's perspective threat model) and also to develop an interactive platform which is able to predict  the consequences of cyberattacks~\cite{2020Datta}. Along this objective, the research team collected the descriptions of a number of cyber security attacks. The selection criterion was based on the Microsoft STRIDE~\cite{STRIDE} threat model by which security threats are classified into six categories: Spoofing, Tampering, Repudiation, Information Disclosure, Denial of Services, and Elevation of Privileges. The research team searched online resources and identified a total number of 102 security attacks
To have a better insight about each attack, three different descriptions for each attack also were  collected from online sources.
The research team then crafted the ``{\it immediate non-technical consequences}'' to end-users for each cyber attack, in which immediate means the consequences and risks that are immediate to end-users. For instance, a cyber attack targeting ``stealing of credentials'' may have the immediate consequences of ``loosing sensitive information,'' but overall consequences of ruining reputation or financial loss for the victim. To avoid any bias in crafting the immediate consequences, a non-computer scientist from the Department of Psychology (the \nth{5} author) read all collected descriptions of selected cyber attacks and developed a short description reflecting the immediate consequences of each attack to end-users. In the end, we developed 50 unique and distinguishable immediate non-technical consequences of the cyber attacks. As an example, Table \ref{tab:EmailSpoofing}  lists the short versions of three different descriptions of ``{\it Email Spoofing}'' searched online along with its immediate consequence to end-user.

\subsection{Group Attacks' Consequences using ``Open Card Sorting''}
\label{sec:group_sorting}

The creation of the 50 unique immediate non-technical consequences of cyber attacks to end-users established the foundation for conducting further research towards creating a usable user-centric security platform where the risk and consequences of each attack is communicated with the users in layman's term (i.e., the short non-technical description of the consequence of each attack presented to average users). However, from usability points of view where each consequences is represented in a unique visual/textual cue, remembering and thus using the 50 different descriptions was considered to be less desirable and infeasible due to the overwhelming cognitive workload of average users. As a result, the research team explored the possibility of reducing the total number of consequences by clustering them into groups. To do so, an ``{\it open card sorting}'' study was designed, a popular technique 
used in human factors to elicit knowledge. 

In a typical card sorting activity, users group concepts into categories with respect to some perceived similarity of the concepts. Furthermore, in a close card sorting the categories are given to users; whereas, in an open card sorting, users are free to create their own categories and group the concepts according to their own perception.

The research team recruited 33 participants including 11 males and 22 females, all introductory psychology students from Texas Tech University, to perform the card-sorting task. The participants did not have any background in computer security or privacy in their work or taken courses. A computer-based system, called Qualtrics\footnote{https://www.qualtrics.com/}, was used to conduct the study and collect the data. The study employed an ``open card sorting'' technique, wherein participants grouped the 50 consequences that they perceived to be similar into participant-created categories. Participants were free to create as many categories (i.e., groups) as they wished, and could organize the consequences into categories as they saw fit based on the similarity of consequences.

The goal was to better understand how end users think about the consequences of security attacks, and to identify a smaller set of consequence categories that could be used as a higher level of  abstraction than dealing with the 50 consequences. To have a feel of card sorting data, Table \ref{tab:cardsorting} shows a snapshot of the participants' card sorting data in which participants 1 and 2 have clustered the 50 consequences into 4 and 3 categories, respectively. A ``1'' on each row indicates that the consequences are grouped together. For example, participant 1 sorted consequences 1 and 3 to be in group \#3.

\begin{table}[t]
\begin{center}
\caption{Three short descriptions of ``{\it Email Spoofing}'' and the immediate non-technical consequence.} 
\begin{tabular}{|p{8cm}|}
\hline
{\bf Description 1.} 
A user receives email that appears to have originated from one source when it actually was sent from another source. Email spoofing is an attempt to trick the user into making a damaging statement or releasing sensitive data~\cite{email1}. \\
{\bf Description 2.} 
The idea of email spoofing is that if an email appears to originate from a known sender, the target of the phishing email will be more likely to fall for the scam ~\cite{email2}. \\
{\bf Description 3.} 
Email spoofing is the forgery of an email header so that the message appears to have originated from someone other than the actual source~\cite{email3}. \\
\hline
{\bf Immediate Consequence.} 
{\it The cyber-attacker made you think that an email that you received from the attacker came from someone else.} \\
\hline
\end{tabular}
\label{tab:EmailSpoofing}
\end{center}
\vspace*{-0.15in}
\end{table}

\begin{table}[t]
\begin{center}
\caption{A snapshot of the consequences grouped through card sorting.} 

\begin{tabular}{|c|c|c|c|c|c|c|c|c|c|c|}
\hline 
\multicolumn{1}{|c|}{\bf Participant} & \multicolumn{1}{|c|}{\bf Group} & \multicolumn{9}{|c|}{\bf 50 Consequences} \\
\cline{3-11}
\# & \# & \multicolumn{1}{|c|}{\bf 1} & \multicolumn{1}{|c|}{\bf 2} & \multicolumn{1}{|c|}{\bf 3} & \multicolumn{1}{|c|}{\bf 4} & \multicolumn{1}{|c|}{\bf 5} & \multicolumn{1}{|c|}{\bf 6} & \multicolumn{1}{|c|}{\bf 7}&\multicolumn{1}{|c|}{\bf 8} & \multicolumn{1}{|c|}{\bf ...}\\
\hline
1 & 1 & 0 & 0 & 0 & 0 & 0 & 0 & 0 & 0 & ... \\
 & 2 & 0 & 1 & 0 & 0 & 0 & 0 & 0 & 0 & ... \\
 & 3 & 1 & 0 & 1 & 0 & 0 & 0 & 0 & 0 & ... \\
 & 4 & 0 & 0 & 0 & 1 & 1 & 1 & 1 & 1 & ... \\
\hline
2 & 1 & 0 & 0 & 0 & 0 & 0 & 0 & 0 & 1 & ... \\
 & 2 & 0 & 0 & 0 & 0 & 1 & 1 & 1 & 0 & ... \\
 & 3 & 1 & 1 & 1 & 0 & 0 & 0 & 0 & 0 & ... \\ 
 \hline
... & ... & ... & ... & ... & ... & ... & ... & ... & ... & ...\\
\hline
\end{tabular}
\label{tab:cardsorting}
\end{center}
\vspace*{-0.2in}
\end{table} 

\subsection{Research Problems and Hypotheses}
\label{sec:group_sorting}

The results of the card sorting study, in which 33 human subjects sorted 50 consequences of cyber threats into participant-generated categories showed a large amount of inter-participant variation.

The research team investigated the participants' data from different perspectives and then developed a set of hypotheses to explain the observed variations: 

\begin{itemize}
\item[--] {\bf {Hypothesis 1 (H1)}}: Each participant employed their own unique way of justification and interpretation of  consequences of cyber threats, explaining the variability of clustering.

\item[--]{\bf {Hypothesis 2 (H2)}}: Due to lack of technical knowledge and carelessness, the participants employed a random-basis clustering when grouping the consequences.

\item[--]{\bf {Hypothesis 3 (H3)}}: The participants might have utilized some logical relationships between textual descriptions of consequences such as (1) frequency-based and (2) semantic-based of terms present in the consequences' description that may explain the participants grouping.

\end{itemize}

{\bf Refuting H1-H2 through Monte Carlo Simulation.} Both H1 and H2 assume a certain degree of randomness. Accordingly, we can evaluate H1 and H2 by determining whether participants group consequences in random ways. To do so, we employed k-means clustering of participants' data (i.e., responses to grouping consequences) to understand how participants grouped consequences. The best clustering solution involved seven categories, and had a Silhouette value of $0.21$, which is still a relatively small value. 

We then executed a Monte Carlo simulation to determine how participants would have grouped the consequences had they done so randomly:
\begin{enumerate}
\item First, the simulation randomly organized the consequences into the same number of groups that each participant created. For example, if a participant created four groups of consequences, the simulation would randomly place the consequences into four groups.
\item  Second, the simulation employed k-means clustering (with seven categories, to match the participant groupings) to determine the Silhouette value. 
\item Third, that process was repeated 100 times. 
\end{enumerate}

Silhouette values based on random groupings ranged from 0.027 to 0.059, which is substantially lower than the observed 0.21. Accordingly, the evaluation suggested participants had not grouped consequences randomly, which ruled out H1 and H2 and shifted our attention to H3. In this paper, we investigate whether hypothesis H3 holds for participants or there might be some other reasons explaining the variations.


\subsection{The Key Novel Idea: Explaining Users through AI}
\label{sec:group_sorting}

This paper focuses on verification of this hypothesis (i.e., H3) by analyzing the textual descriptions of attack consequences using Natural Language Processing (NLP) techniques. The NLP techniques explored involved 
\begin{itemize}
    \item[--] Frequency-based (e.g., Bag of Words (BoW), Term Frequency – Inverse Document Frequency (TF-IDF), and Latent Semantic analysis (LSA)) where the tokens of given documents are considered independent than the context and semantic, and
    \item[--]  Semantic-based (e.g., WordNet) where the context of each word in the given documents is considered.  
\end{itemize}

The goal is to explore whether it is possible to understand participants' perception through NLP analysis. Accordingly, we build models based on these NLP techniques using the textual descriptions of attacks' immediate and non-technical consequences and then compare their outputs with the participants' card sorting data. A closer similarity between the participants' data and certain outputs produced by NLP techniques is an indication that the participants have utilized similar strategy to perceive the consequences of security attacks. Therefore, a high similarity of the users' card sorting data with frequency-based NLP shows that the participants utilized each keyword in the description of consequences without paying attention to the semantic of each consequence; whereas, a high similarity of the users' card sorting data with semantic-based approaches is an indication that the users were aware of the meaning and semantic of each consequences.

The deriving idea of this research is that if the results of the NLP-based algorithms are comparable with the participants data, it would shed some lights on how the participants thought (i.e., participants perception) when grouping the consequences of cyber attacks. As a result of this research, we propose a new line of research through which users' mental models and perceptions are explained through AI in general and using NLP in specific. The term ``{\it Explainable Users}'' then is coined to demonstrate understanding users' behavior by the help of AI techniques.  

\subsection{The Key Contributions}
\label{sec:group_sorting}

The key contributions of this paper are as follows:
\begin{enumerate}
\item  We introduce the novel idea of ``{\it Explainable Users}'' where AI techniques are employed to interpret users' mental model, behavior, and perceptions. 
\item We introduce the idea of application of NLP-based techniques (i.s., a branch of AI) for understanding the user's mindset and their perceptions towards an application domain in general and the cyber security context in specific. 
\item The results show that average users with little to no security  knowledge pick only individual terms appeared in the textual description of cyber attacks when comprehending their consequences. 
\item The results indicate that, due to limited cyber security background, the perception of average users are less driven by the semantic of cyber attacks.  

\end{enumerate}

The rest of this paper is organized as follows: Section \ref{sec:relatedwork} briefly reviews the related work. Section \ref{sec:conducting} justifies the experiments in this paper and shows the methodology followed. Section \ref{sec:alg} presents the theoretical background for the NLP-based algorithms studied in this paper. Section \ref{sec:form_sims} shows how the similarity matrices were generated for each NLP-based algorithm's output. The implementation details are provided in Section \ref{sec:Imp}. Section \ref{sec:results} presents the results and discussions and finally Section \ref{sec:conclusion} concludes the paper.

\section{Related Work}
\label{sec:relatedwork}

There exist some studies about investigation of users' mental model exposing cyber security textual data~\cite{2007Asgharpour,2008Camp,2020Cotoranu}.These studies mostly focus on how expert users are different from novices analyzing cyber security textual data. Similar applications of NLP-based algorithms are reported for  predicting user's click behavior~\cite{KaurH05}. However, to the best of our knowledge, such an NLP-based approach to model users' mindset when comprehending the consequences of security attacks has not been discussed in the literature.

Robertson et al.~\cite{2020Robertson} conducted both closed and open card sorting experiments to compare novices and experts’ performance in the domain of ``trust in self-driving vehicles.'' 
In the open card sorting, the participants were asked to group the items and then label the groups; Whereas, in the closed card sorting, the participants were asked to group items within pre-determined labels. They computed the Jaccard/Tanimoto coefficients to measure the similarity of four card sorting experiments including 1) Expert Open vs. Novice Open, 2) Expert Open vs. Novice Closed, 3) Expert Close vs. Novice Open, and 4) Expert Close vs. Novice Closed. They obtained the highest average similarity for Expert Close vs. Novice Closed and the lowest average similarity for Expert Open vs. Novice Open. The authors concluded that novices and experts in closed card sort are performing more alike compared to open card sorting.

With the goal of higher quality of security risk communication, Asgharpour et al.~\cite{2007Asgharpour} conducted a closed card sorting experiment on security risk words to investigate the mental model of novices and experts. 
They formed similarity matrices for experts and non-experts. Using those matrices, they computed the word distance matrices for both teams. They utilized Multidimensional Scaling (MDS) technique to map distance matrices to a 2D space in order to find structure in the set of distances and compare the mindset of two teams. Their results showed that for 10 out of 29 risks, experts and non-experts exhibit different mental models.

Cotoranu and Chen~\cite{2020Cotoranu} performed text analysis to investigate the mental model of learners with two different levels of familiarity with cyber security subjects.
They recruited undergraduate students divided into two groups and asked them to answer three open-ended cyber security questions.
They extracted bigrams from the preprocessed data. First, they computed the total frequency of bigrams for each group. The results showed that students with higher level of knowledge about cyber security write more. Second, they calculated the entropy to measure the lexical variation for both groups. The results showed that students with higher level of knowledge about cyber security tend to use more domain specific words and hence write sentences that are more complex. Third, they used cosine similarity to compare the answers for these two groups. The results indicated the gap between the mental models for these two groups.

This paper introduces the novel idea of utilizing AI and more specifically NLP techniques for analyzing mental model and perceptions of novice users when dealing with textual description of consequences of security attacks. This idea creates a new line of research in modeling human factors using AI techniques. 

\section{Discovering users' perception through NLP}
\label{sec:conducting}

Considering the fact that the participants clustered consequences using the given textual descriptions, the research team wondered whether it was possible to employ NLP-based techniques to analyze the textual consequences and then compare them with the participants data and investigate whether or not they are exhibiting any similarities. 

More specifically, the objective was to discover if there was any similarity between the participants' card sorting data and the outputs of these NLP-based algorithms in order to make any judgment about the participants' thoughts and the way the immediate consequences were grouped. There exist a good number of NLP-based techniques and tools to analyze data whose features are extracted using different approaches as a basis (e.g., semantic-based, frequency-based, etc.). These algorithms can serve as a baseline for comparing data obtained from participants. 

To address the research problem stated above, we implemented four NLP techniques 1) Bag of Words (BoW), 2) Term Frequency-Inverse Document Frequency (TF-IDF), 3) Latent Semantic Analysis (LSA), and  4) WordNet to analyze textual data and compare it to the data sorted by the participants. 

In order to investigate to what extent participants’ consequence sorting result is similar to different NLP-based approaches, we decided to form a similarity matrix for each NLP-based approach that contains pairwise similarity values for the consequences. To this end, we first extracted the features from consequences using each NLP-based approach and then computed the similarity values for each pair of features. 
Moreover, we explored the Pearson correlations between the participants' similarity data and the output of various  NLP-based algorithms. We applied BoW, TF-IDF, and LSA  while taking into account different preprocessings of the text regarding 1) the presence of stop words (i.e., words that do not offer content and serve only for syntactic purposes), 2) stemming (i.e., extracting the root of the words), and also 3) considering $n$-grams (i.e., subsequences of $n$ words) where $n = 1, 2, 3$. 

\subsection{The Characteristics of NLP Techniques for Perceptions}

There are some intriguing reasons that explain why these NLP-based techniques might be able to reveal users' perceptions when the underlying task is merely reasoning based on textual descriptions. Although more technical details are presented in Section \ref{sec:alg}, here we discuss about how these techniques are related to users' perception. 

{\bf Bag of Words (BoW)}. This method assigns more weight to the words (i.e., tokens) that are repeated more in the corpus (i.e., the set of all consequences). As a result, higher correlation between BoW and participants' data indicates that the participants were more considering the words (tokens) that are repeated more in the set of all consequences 
when they were sorting consequences.



{\bf Term Frequency–Inverse Document Frequency (TF-IDF)}.
In this method, the weight related to each word (token) increases with its repetition in the sentence and decreases with its repetition in the whole document (i.e., the set of all consequences). As a result, this method assigns more weight to words that are repeated uniquely in the sentence and not in the other documents (i.e., all consequences). Therefore, higher correlation between TF-IDF and participants' data is an indication that the participants were comparing each sentence with other sentences in other consequences to find the repeated words (i.e., tokens) that are more unique in that sentence when they were sorting consequences.



{\bf Latent Semantic Analysis (LSA)}. This method enables analyzing the use of words and groups of words in texts. In LSA, the set of words (i.e., tokens) extracted by TF-IDF is reduced to a certain number. In LSA, features (i.e., words and their weights) are transformed with the goal of finding the hidden topics through analyzing the ``{\it context''} of the given text. Hence, LSA enables understanding the meaning of the text and the effects of words on the meaning of documents. It also enables us to study the average meaning of words in a document in accordance with the overall meaning the document.



{\bf WordNet}. The technique used in WordNet works based on considering the synonyms and grouping of words with similar concepts. As a result, higher correlation between WordNet and participants' data indicates that the participants were beyond only considering the syntactical repeated words (tokens) superficially and had employed a deeper, synonym-based and semantical analysis when they were sorting consequences.



\section{Technical Details of NLP-based Algorithms}
\label{sec:alg}

To investigate the research hypothesis H3, the research team implemented a number of Natural Language Processing (NLP) algorithms including 1) Bag of Words (BoW), 2) Term Frequency-Inverse Document Frequency (TF-IDF), 3) Latent Semantic Analysis (LSA), and 4) WordNet.

\subsection{BoW}
The Bag of Words (BoW) is a simple and flexible approach for feature extraction from text. It can be considered as a histogram of words (or tokens) within the document along with the number of their occurrence within the text~\cite{2017Goldberg}. The word ``bag'' indicates that the order and structure of the words (or tokens) are not considered. The bag of words depends on the vocabulary of the text. So, as the number of sentences increases, the vocabulary would increase exponentially. Hence, bag of words works better when the data is context-specific.

\subsection{TF-IDF}


The TF-IDF algorithm has several applications including the identification of important terms, measuring the rareness of terms, and categorization of documents. 
The TF-IDF weight is composed of two parts: 

{\bf TF}: The normalized Term Frequency (TF) is the number of times a word appears in a document divided by the total number of words in that document. More specifically, TF measures how ``{\it frequently}'' a term occurs in a document. Since every document is different in length, it is possible that a term would appear many times in long documents than the shorter ones. Therefore, the term frequency is often divided (i.e., normalized) by the document length, i.e., the total number of terms in the document, as a way of normalization: 

\vspace{-0.12in}
\begin{equation*}
TF(t) = \frac{(\#\ term\ t\ appears\ in\ a\ document)} {(\#\  number\ of\ terms\ in\ the\ document)}
\end{equation*}

{\bf IDF}: The Inverse Document Frequency (IDF) is the logarithm of the number of the documents in the corpus divided by the number of documents where the specific term appears. More specifically, IDF measures how important a term is. While computing TF, all terms are considered equally important;  it is known that certain terms, such as ``is'', ``of'', and ``that'' may appear a lot of times but have little importance. Thus, we need to weigh down the frequent terms while scale up the rare ones, by computing the following: 

\vspace{-0.12in}
\begin{equation*}
IDF(t) = log(\frac{\# documents}{\# documents\ with\ term\ t})
\end{equation*}

The magnitude of the weight value computed for each term shows  the frequency of the term and thus indicates how common/rare the term is in the entire set of collections. 

\subsection{LSA}

Latent semantic analysis (LSA) is used for analyzing the relationships between a set of documents and the terms they contain by producing a set of concepts related to the documents and terms. The basic assumption, also called distributional hypothesis, is that words that are close in meaning will occur in similar parts of text. 
LSA builds a matrix containing word counts per paragraph from a large corpus. Then the Singular Value Decomposition (SVD), a very common technique used in variable reductions including Principal Component Analysis (PCA), is applied in order to reduce the number of rows while maintaining the similarity and structure of columns. 
LSA is a method for extracting and representing the contextual-usage meaning of words by statistical computations applied to a large corpus of text~\cite{LSA}. The deriving idea is that the aggregated set of word contexts in which a given word does and does not appear determines the similarity of meaning of words. 

LSA has several important applications in NLP. More notably, it mimics human word sorting and category judgments, simulates word-word and passage-word lexical data, estimates passage coherence, measures the learnability of passages, and enables assessing the quality of knowledge in an essay~\cite{LSA}.

\subsection{WordNet}

WordNet~\cite{WordNet} is a large lexical database of English words representing part-of-speech such as nouns, verbs, adjectives and adverbs which are grouped into sets of ``{\it cognitive synonyms}'', also called synsets. Each synset expresses a distinct concept that are interlinked through conceptual-semantic and lexical relations creating a network of conceptually meaningful and interrelated words. 
Synonym of words is the main building block in WordNet. Accordingly, the synsets in WordNet are linked together through ``conceptual relations.'' For instance, a general synset like \{furniture\} is linked to a more specific synset such as \{bed\} and \{bunkbed\}. More specifically, WordNet builds up a hierarchy in which the category 
\{furniture\} includes the category \{bed\}, which in turn includes the category \{bunkbed\}. In an analogous way, concepts such as \{bed\} and \{bunkbed\} are building up the category \{furniture\}.  

WordNet also enables expressing ``part-whole'' relations between synsets in which a composition relation can be defined (e.g.,  synsets such as \{car\} and \{tires\} and \{engine\}). Moreover, inheritance relationships, also called ``super ordinates'' can be also considered in WordNet. WordNet mostly connects words from the same part of speech (PoS) consisting of four subnets  representing nouns, verbs, adjectives, and adverbs.

\section{Forming Similarity Matrices of Consequences}
\label{sec:form_sims}

The output of the count-based NLP algorithms (i.e., BoW, TF-IDF, and LSA) is a matrix representation of the consequences which will be used to calculate the pairwise similarities between them. 

The similarity matrices for the BoW, TF-IDF, and LSA representations were calculated using the cosine similarity for each pair of consequences. This is defined as

\vspace{-0.1in}
\[
    CosSim(c_1, c_2) = \frac{c_1 \cdot c_2}{||c_1|| ||c_2||},
\]

\noindent for rows of the matrix representation belonging to consequences $c_1$ and $c_2$.

Unlike the matrices for the other NLP-based algorithms, the WordNet similarity matrix was calculated using the Greedy Lemma Aligning Overlap~\cite{vsaric2012takelab}, defined as

\vspace{-0.1in}
\[
Sim(S_1, S_2) = \frac{\sum_{(l_1, l_2) \in P} Sim(l_1, l_2)}{max(length(S_1), length(S_2))}
\]

\noindent where $P$ is the Cartesian product between the synsets for the words in $S_1$ and $S_2$,  $Sim(l_1, l_2)$ is the path similarity between words $l_1$ and $l_2$, which corresponds to the shortest path that connects the two synsets for $l_1$ and $l_2$ in the hierarchical structure of WordNet. Also, this metric is normalized by the length of the longer sentence. Note that this approach does not need an intermediate matrix representation for consequences.

Finally, the similarity matrix for the data provided by the participants was calculated. The consequences data clustered by all participants was treated as a set of observations for each column, i.e., a consequence. The Spearman's rank correlation value between two columns (i.e., consequences) in Table \ref{tab:cardsorting} represents how two different consequences are similar in the participants' card sorting data. Although the similarity matrices were fully generated, we only used their triangular matrices ignoring the diagonal when calculating the Pearson correlation with the participants' similarities. This is performed to avoid redundant information in the correlations.

\section{Implementation}
\label{sec:Imp}

We developed several Python 3 scripts for performing our study. For implementing stemming, we used the Porter stemming algorithm provided in the NLTK library~\cite{loper2002nltk}, whereas we used the SpaCy library~\cite{spacy2} for lemmatization. Lemminization was essential when capturing the similarities through WordNet. Because of the nature of the WordNet algorithm, considering $n$-grams and performing stemming could not be possible. Instead, we applied the WordNet algorithm considering the presence of stopwords and lemmatization. Although both stemming and lemmatization achieve the same results, which is text normalization, lemmatization ensures that the roots of the words are valid English words.

\section{Results and Discussion}
\label{sec:results}

\subsection{Correlation Comparison of NLP Techniques}

\begin{figure}[t!]
\vspace*{-0.2in}
\centering
\includegraphics[width=0.25\textwidth]{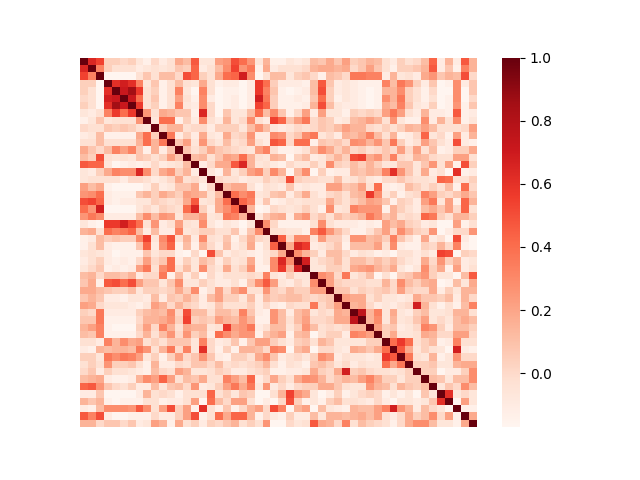}
\caption{Similarity heatmap for pairwise consequences according to the participants' data.}
\label{fig:participantdata}
\vspace*{-0.1in}
\end{figure}

Figure \ref{fig:participantdata} shows the heatmap of the pairwise consequence similarity matrix for the participants' data. Onward, this matrix will be used as the ground truth when calculating its correlation with the outputs of the NLP-based algorithms.
Table \ref{tab:all_corr} shows the Pearson correlation between the user's similarity matrix and the similarity matrices generated by BoW, TF-IDF, and LSA for each configuration considering the status of stop words, stemming, and $n$-grams. In addition, we group the correlations according to the status of stop words and stemming.
As to the $n$-grams, Table \ref{tab:all_corr} reports the correlations between each technique and the participants' data using unigrams (Uni), bigrams (Bi), and trigrams (Tri).

\begin{table}[!h]
\centering
    \caption{Pearson correlation between NLP-based and user-based similarities under different scenarios.}
    \label{tab:all_corr}
    \centering
    \begin{tabular}{|c|c|c|c|l|l|l|}
    \hline
    \multicolumn{4}{|c|}{\bf Configuration} & \multicolumn{3}{c|}{\bf Pearson Correlation} \\ 
    \hline
     & {\it Stop Words} & {\it Stemming} & {\it N-gram} & {\it BoW} & {\it TF-IDF} & {\it LSA} \\ 
     & {\it Included?} & {\it Applied?} & {\it } & {\it } & {\it } & {\it } \\ \hline
    1 & \textbf{No} & \textbf{No} & \textbf{Uni} & \textbf{0.602} & \textbf{0.556} & \textbf{0.522} \\ \hline
     & No & No & Bi & 0.450 & 0.445 & 0.284 \\ \hline
     & No & No & Tri & 0.394 & 0.385 & 0.176 \\ \hline
    \hline
    2 & No & Yes & Uni & 0.555 & 0.518 & 0.453 \\ \hline
    & No & Yes & Bi & 0.455 & 0.450 & 0.289 \\ \hline
    & No & Yes & Tri & 0.403 & 0.396 & 0.179 \\ \hline
    \hline
    3 & Yes & No & Uni & 0.517 & 0.495 & 0.418 \\ \hline
    & Yes & No & Bi & 0.528 & 0.484 & 0.361 \\ \hline
    & Yes & No & Tri & 0.530 & 0.416 & 0.258 \\ \hline
    \hline
    4 & Yes & Yes & Uni & 0.498 & 0.477 & 0.405 \\ \hline
     & Yes & Yes & Bi & 0.497 & 0.471 & 0.324 \\ \hline
    & Yes & Yes & Tri & 0.433 & 0.419 & 0.268 \\ \hline
    \end{tabular}%
    
\end{table}


\begin{table}[!h]
    \caption{Pearson correlation between the WordNet-based and user-based similarities.}
    \label{tab:wn_corr}
    \centering
    \begin{tabular}{|c|c|c|}
    \hline
    \multicolumn{2}{|c|}{\bf WordNet} & {\bf Pearson Correlation} \\ \cline{1-2}
    {\it Stop Words Included?} & {\it Lemma Applied? } & \multicolumn{1}{c|}{} \\ \hline
    \textbf{No} & \textbf{No} & \textbf{0.425} \\ \hline
    No & Yes & 0.373 \\ \hline
    Yes & No & 0.366 \\ \hline
    Yes & Yes & 0.299 \\ \hline
    \end{tabular}%
   
\end{table}

Table \ref{tab:wn_corr} shows the Pearson correlation between the user's similarity matrix and the output matrices of the WordNet algorithm under the conditions described in Section \ref{sec:conducting}. 
One commonality across all the NLP-based techniques is that the highest Pearson correlation is achieved using text without stop words, using unigrams (for BoW, TF-IDF, and LSA), and without text normalization (stemming for BoW, TFDF, and LSA; lemmatization for WordNet). The highest correlations are 0.602, 0.556, 0.522, and 0.425 for BoW, TF-IDF, LSA, and WordNet, respectively.

\begin{figure*}[t!]
\vspace*{-0.02in}
    \begin{subfigure}{0.25\textwidth}
      \includegraphics[width=\textwidth]{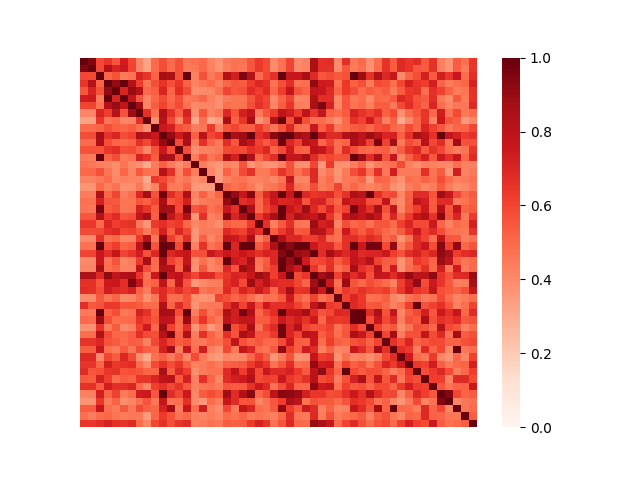}
     
      \caption{WordNet}
      \label{fig:heat_wn}
    \end{subfigure}%
    \begin{subfigure}{0.25\textwidth}
      \includegraphics[width=\textwidth]{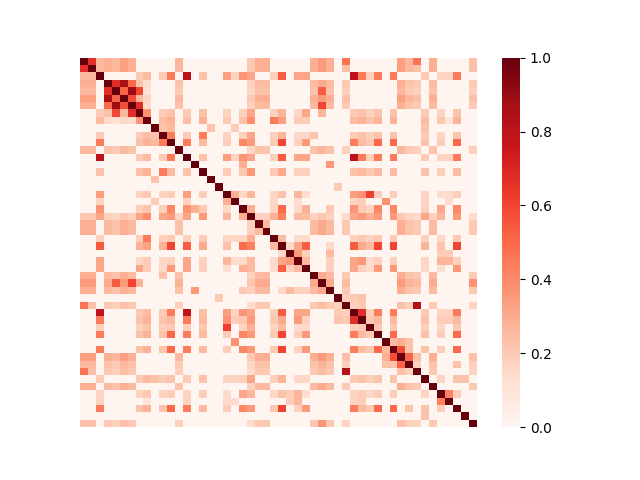}
   
      \caption{BoW}
      \label{fig:heat_bow}
    \end{subfigure}%
    \begin{subfigure}{0.25\textwidth}
      \includegraphics[width=\textwidth]{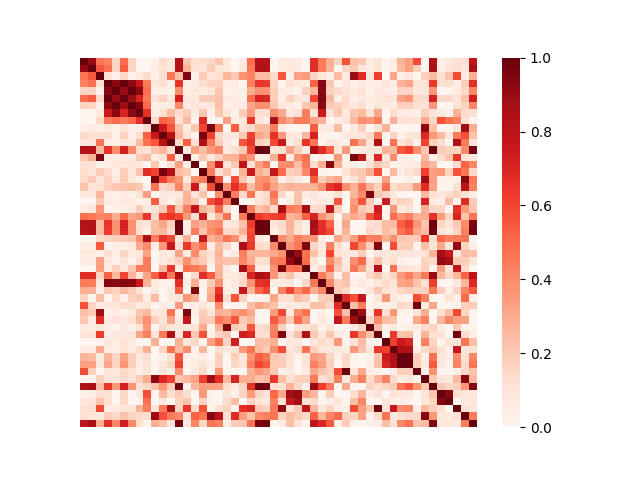}
      
      \caption{LSA}
      \label{fig:heat_lsa}
    \end{subfigure}%
    \begin{subfigure}{0.25\textwidth}
      \includegraphics[width=\textwidth]{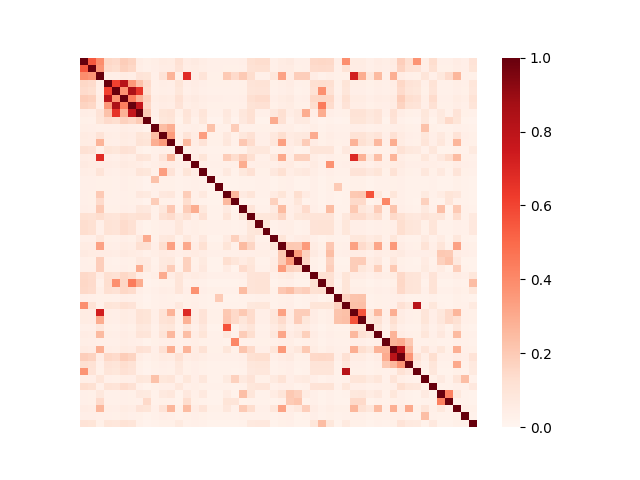}
      
      \caption{TF-IDF}
      \label{fig:heat_tfidf}
    \end{subfigure}%
    \caption{The heatmaps  for  the  pairwise consequences similarity  matrices of the NLP-based algorithms with the highest Pearson correlation with participants' data. The highest Pearson correlation with participants' data for \ref{fig:heat_bow},\ref{fig:heat_lsa},\ref{fig:heat_tfidf} achieved by ``not including stop words", ``not applying stemming" and ``considering unigrams". The highest Pearson correlation with participants' data for \ref{fig:heat_wn} achieved by ``not including stop words" and ``not applying lemmatization".
    }
    \label{fig:heatmaps}
    \vspace*{-0.1in}
\end{figure*}


Figure \ref{fig:heatmaps} shows the heatmaps of the pairwise similarity of consequences for the NLP techniques implemented. Note that we report the Pearson Correlation under several configurations of our experiments (See Table \ref{tab:all_corr} and \ref{tab:wn_corr}), because of this, the heatmaps in Figure \ref{fig:heatmaps} are those obtained using the configurations that reported the highest Pearson correlation (i.e., first rows of Table \ref{tab:all_corr} and \ref{tab:wn_corr}). 

Figure \ref{fig:heat_wn} presents higher similarity values when compared to the rest of heatmaps, this is due to the lack of an intermediate sparse and high-dimensional matrix representation when applying the WordNet-based algorithm. However, note that consistently higher similarities between consequences in WordNet did not correlate as high as the frequency-based approaches. In fact, the maximum Pearson correlation in Table \ref{tab:wn_corr} (i.e., WordNet-based technique) is 0.425, whereas the highest correlation in Table \ref{tab:all_corr} (i.e., frequency-based techniques) is 0.602 for BoW, approximately 41\% higher than the highest correlation in Table \ref{tab:wn_corr}. Also, note that the highest correlation in Table \ref{tab:all_corr} (BoW) accounts for the 36.24\% of the explained variance (i.e., coefficient of determination $R^2$), in contrast to the Table \ref{tab:wn_corr}'s (WordNet) 18.06\%. This finding suggests that the participants performed the consequence-sorting using information taken directly from the text rather than considering further semantics of the terms.

One effect of the dimensionality reduction in LSA can also be seen in Figure \ref{fig:heat_lsa}, where the less sparse resulting matrix of LSA produces higher similarity values between consequences when compared to BoW and TF-IDF. Likewise, the results of WordNet, higher similarity values computed for consequences did not report higher correlation values between participants' data and the output of LSA among the frequency-based approaches. 
Moreover, the fact that the best correlations between participants' data and frequency-based approaches are consistently higher than that of the WordNet algorithm agrees partially with {\bf H3}.
A reason for this specific behavior could be the participants' limited or lack of enough exposure to cyber security terminology. This would cause the sorting of consequences performed purely with the presented text as a basis and not allowing a deeper semantic consideration, which is what the WordNet knowledge-base offers with its hierarchy.
Furthermore, for all frequency-based algorithms unigrams report the highest correlations with participants' data. This suggests that participants where grouping consequences according to the occurrences of words independently, and not taking into account co-occurrences of words in a small context window, as feature extraction using bigrams/trigrams operates.

Finally, the dominance of BoW correlation values with participants' data over TF-IDF in all peer configuration indicates that participants considered the importance of keywords in each consequence separately and did not considered if that key word is repeated in all the consequences. This result derives from the fact that TF-IDF offers more weight to words, which appear in fewer sentences and decreases the weight for a word as it appears in more sentences (i.e., consequences).

To sum up, according to the findings above, the research team describes participants mental model for consequences card sorting as follows: participants clustered cyber attack consequences based on the single keywords  in each sentence and less considered the semantic behind the sentences. The results support the idea that participants in this experiment used shallow analysis for this card sorting study. This result is consistent with previous research that novice users do not deal with deep cues in an open card sorting problem~\cite{2020Robertson}.

{\bf Psychological Perspective.} From psychological point of view, the present results could reflect that the participants were non-experts. Non-experts think differently than experts about cybersecurity~\cite{2007Asgharpour,2013Bartsch,2011Bravo,2017Theofanos}. For example, non-experts think about cybersecurity in more abstract ways than experts~\cite{2013Bartsch}, are less likely than experts to think about topics such as information security~\cite{2011Bravo} or risk factors and consequences of threats~\cite{2013Bartsch}, are less likely than experts to think they can protect themselves~\cite{2013Bartsch,2017Theofanos}, are more likely than experts to think websites can be trusted to protect users’ cybersecurity~\cite{2017Theofanos}, and are more likely than experts to think about whether a website looks professional when deciding whether it is trustworthy~\cite{2011Bravo}. Accordingly, the participants in the present study may have focused more on key terms in the consequence descriptions and less on semantics or context because they likely lacked a deep understanding of cybersecurity issues. Future research could test this possibility by replicating the present research but with cybersecurity experts.

\subsection{Sensitivity Analysis}

An interesting research question is how the different combinations of stop words inclusion and stemming affects the correlation values between participants' data and the output of each NLP technique. For instance, if the participants pay attention to the roots of the words when sorting the consequences' descriptions, then the Pearson correlations in Table \ref{tab:all_corr} should present major differences depending on whether the stemming preprocessing step was performed and other elements of the configuration are fixed (i.e., inclusion of stop words and N-gram). Consider the highest correlation values reported in Table \ref{tab:all_corr} for the first group (stop words not included, stemming not applied) and the second group (stop words not included and stemming applied), the Pearson correlation values between participants' data and BoW are 0.602 and 0.555, respectively. In this case, not performing stemming while removing the stop words in both cases increases the correlation value by 8.46\%.

On the other hand, and still for BoW, keeping the text not-stemmed while including (third group) or removing (first group) stop words reports a 16.4\% increase of correlation for the non-including stop words case. This suggests that the participants are more attentive to content words (e.g., keywords), whereas they did not consider the roots of the words at the same extent. 

Finally, consider the three correlations in the first group of BoW, in this case both stop words and stemming conditions are fixed to not included and not applied, respectively. Under this configuration, the use of unigrams reports an increase of 52.79\% and 33.77\% in the correlation when compared to trigrams and bigrams, respectively. This implies that the participants were looking for occurrences of words individually and not in pairs or triples.

\begin{figure}[!t]
\vspace*{-0.3in}
\centering
\includegraphics[width=0.80\linewidth]{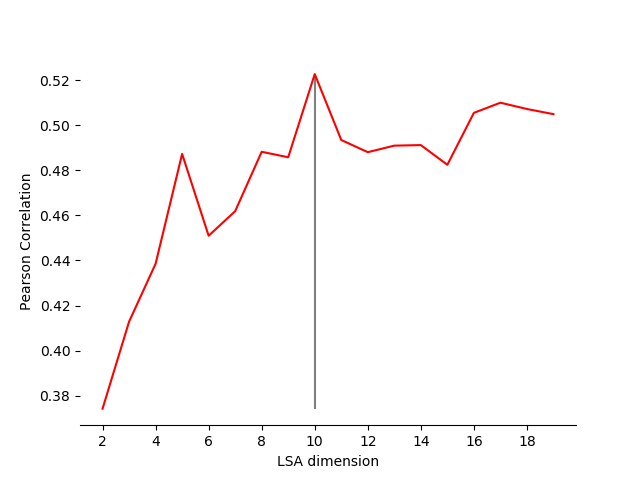}
\caption{Pearson correlation for LSA and participants' data considering different dimensions for LSA.}
\label{fig:lsadimensions}
\vspace*{-0.1in}
\end{figure}

\subsection{The Effect of LSA Dimensions on Correlations}
One important aspect of LSA is the number of dimensions of the resulting matrix after applying SVD, which has to be determined beforehand. In this work, we tried different number of dimensions ranging from 2 to 20 and selected the one that yielded the highest Pearson correlation with the participants' data. With this criterion, as seen in Figure \ref{fig:lsadimensions}, the optimum number of dimensions was 10 and hence, we considered this value for computing correlation values for LSA and participants' data. It should be mentioned that since the correlation value for TF-IDF is higher than LSA it shows, the highest correlation can be obtained when SVD is not applied.

\section{Conclusion and Future work}
\label{sec:conclusion}
This paper introduces the idea of utilizing NLP-based approaches to understand user's thoughts and perception on cyber-attacks. We implemented three frequency-based NLP techniques (BoW, TF-IDF, and LSA), and one semantic-based (WordNet). Using these techniques, we extracted the features for each one and then calculated the similarity between each pair of consequences using each set of features separately. Finally we compared the pairwise similarities obtained from each technique with pairwise similarities obtained from participants' card sorting.

We acquired several general findings from this experiment. First, we conclude that participants sorted the consequences based on analyzing the keywords for each consequence. This is due to the highest correlations obtained from removing stop words in all applied NLP techniques. Second, we can infer that participants sorted the consequences by considering the repetition of similar keywords. This is because of the dominance of the correlation values from frequency-based NLP approaches over semantic-based NLP approach, especially considering the top-score correlation for BoW. Third, it was shown that participants interpret similar consequences by analyzing the sentences word by word. This result was concluded from the fact that the unigrams property of all frequency-based approaches resulted in the highest correlation values. Finally, The low correlation values for WordNet technique (less than 0.5) reveals the fact that participants did not considered the semantics of the sentences related to each consequence. 

The proposed approach can be used as a baseline for further investigation. In this work, we calculated the correlation between similarities considering the grouping of consequences of the 33 participants at once. However, it is also known that the inter-participant variance in the sorting is high; thus, a possibly better approach would be to cluster the participants themselves according to how they performed the sorting task (e.g., by the number of groups created by each participant, how the consequences were sorted, or a mix of both) and next apply the same NLP techniques or utilize document embedding methods such as Doc2Vec~\cite{2020Gutierrez} to each resulting cluster of participants. 

It is shown that generally the difference between novices and experts in an open card sorting is higher than a closed open card sorting~\cite{2020Robertson}. Considering such a study and current work, another informative experiment is to perform the card-sorting process for consequences by expert people in the cyber security area and then conduct the same analysis on the provided data. With both results, we can compare the data from novice users against expert users and study how these two groups conceptualize consequences of cyber attacks.


\section*{Acknowledgement}

This research work is supported by National Science Foundation (NSF) under Grant No. 1564293.

\bibliographystyle{IEEEtran}
\bibliography{IEEEfull,sample-bibliography}

\end{document}